\documentclass[
    prl, twocolumn, superscriptaddress, amsmath, amssymb,
    aps, floatfix, preprintnumbers
]{revtex4-2}
\usepackage[utf8]{inputenc}
\usepackage{setspace}
\usepackage[dvips]{graphicx}
\usepackage{booktabs}
\usepackage[dvipsnames]{xcolor}
\usepackage{tikz}
\definecolor{CiteBlue}{RGB}{45,52,151}
\usepackage[
    colorlinks=true,
    linkcolor=CiteBlue,
    urlcolor=CiteBlue,
    citecolor=CiteBlue
]{hyperref}
\usepackage[version=4]{mhchem}
\usepackage{aas_macros}

\usepackage[capitalise]{cleveref}
\usepackage{siunitx}
\DeclareSIUnit{\year}{yr}

\let\oldsection\section
\usepackage{multibib}
\newcites{sm}{SM references}

\newcommand{\refcite}[1]{Ref.~\cite{#1}}
\newcommand{\refscite}[1]{Refs.~\cite{#1}}
\newcommand{\refcitesm}[1]{Ref.~\citesm{#1}}
\newcommand{\refscitesm}[1]{Refs.~\citesm{#1}}

\usepackage{bm}
\newcommand{\bb}[1]{\bm{\mathrm{#1}}}
\newcommand{\du}{\mathrm{d}}
\newcommand{\dd}{\,\du}
\newcommand{\dm}{\mathrm{DM}}
\newcommand{\med}{\phi}
\renewcommand{\Im}{\operatorname{Im}}
\renewcommand{\Re}{\operatorname{Re}}

\newcommand{\el}{\mathrm{e}}
\newcommand{\lind}{\mathrm{L}}
\newcommand{\fermi}{\mathrm{F}}

\newlength{\twosubht}
\newsavebox{\twosubbox}

\begin{document}

\title{First Sub-MeV Dark Matter Search with the QROCODILE Experiment Using Superconducting Nanowire Single-Photon Detectors}
\preprint{MIT-CTP/5744}

\author{Laura Baudis}
\author{Alexander Bismark}
\author{Noah Brugger}
\author{Chiara Capelli}
\author{Ilya Charaev}
\author{Jose Cuenca Garc\'\i{}a}
\affiliation{Department of Physics, University of Z\"urich, Winterthurerstrasse 190, CH-8057 Z\"urich, Switzerland}

\author{Guy Daniel Hadas}
\affiliation{Racah Institute of Physics, Hebrew University of Jerusalem, Jerusalem 91904, Israel}
\author{Yonit Hochberg}
\affiliation{Racah Institute of Physics, Hebrew University of Jerusalem, Jerusalem 91904, Israel}
\affiliation{Laboratory for Elementary Particle Physics,
 Cornell University, Ithaca, NY 14853, USA}

\author{Judith K. Hohmann}
\affiliation{Karlsruhe Nano Micro Facility (KNMFi), Engesserstrasse 5, 76131 
Karlsruhe, Germany}
\affiliation{Karlsruhe Institute of Technology, Engesserstrasse 5, 76131 
Karlsruhe, Germany}

\author{Alexander Kavner}
\affiliation{Department of Physics, University of Z\"urich, Winterthurerstrasse 190, CH-8057 Z\"urich, Switzerland}

\author{Christian Koos}
\affiliation{Karlsruhe Institute of Technology, Engesserstrasse 5, 76131 
Karlsruhe, Germany}

\author{Artem Kuzmin}
\affiliation{Karlsruhe Institute of Technology, Engesserstrasse 5, 76131 
Karlsruhe, Germany}

\author{Benjamin V. Lehmann}
\affiliation{Center for Theoretical Physics, Massachusetts Institute of Technology, Cambridge, MA 02139, USA}

\author{Severin N\"ageli}
\author{Titus Neupert}
\author{Bjoern Penning}
\author{Diego Ram\'\i{}rez Garc\'\i{}a}
\author{Andreas Schilling}
\affiliation{Department of Physics, University of Z\"urich, Winterthurerstrasse 190, CH-8057 Z\"urich, Switzerland}


\date\today

\begin{abstract}\ignorespaces{}
    We present the first results from the \textit{Quantum Resolution-Optimized Cryogenic Observatory for Dark matter Incident at Low Energy} (QROCODILE). The QROCODILE experiment uses a microwire-based superconducting nanowire single-photon detector (SNSPD) as a target and sensor for dark matter scattering and absorption, and is sensitive to energy deposits as low as \qty{0.11}{\electronvolt}. We introduce the experimental configuration and report new world-leading constraints on the interactions of sub-MeV dark matter particles with masses as low as \qty{30}{\kilo\electronvolt}. The thin-layer geometry of the system provides anisotropy in the interaction rate, enabling directional sensitivity. In addition, we leverage the coupling between phonons and quasiparticles in the detector to simultaneously constrain interactions with both electrons and nucleons. We discuss the potential for improvements to both the energy threshold and effective volume of the experiment in the coming years.
\end{abstract}

\maketitle

\section{Introduction}
\label{sec:intro}
Laboratory searches for dark matter (DM) particles have played a key role in constraining DM candidates at the weak scale~\cite{ParticleDataGroup:2024cfk,deSwart:2017heh,Lee:1977ua,Kolb:1990vq,Jungman:1995df,Bergstrom:2000pn,Bertone:2004pz}. More recently, direct detection experiments have begun to probe a new frontier: light DM, with mass well below the weak scale~\cite{Essig:2011nj,Graham:2012su,Essig:2015cda,Hochberg:2015pha,Hochberg:2015fth,Hochberg:2019cyy,Hochberg:2021ymx,Hochberg:2021yud,Derenzo:2016fse,Hochberg:2016ntt,Hochberg:2017wce,Cavoto:2017otc,Kurinsky:2019pgb,Blanco:2019lrf,Griffin:2020lgd,Simchony:2024kcn,Essig:2022dfa,Das:2022srn,Das:2024jdz}. DM interactions are relatively unconstrained at masses between the keV and GeV scales, and many new experiments have been devised to search for DM in this regime~\cite{Essig:2022dfa,SENSEI:2023zdf,Hochberg:2019cyy,Hochberg:2021yud,Gao:2024irf}. Direct detection for DM masses below $\mathord{\sim}\qty{1}{\giga\electronvolt}$ faces considerable challenges. Such DM particles are lighter than atomic nuclei, limiting the energy that can be transferred in a free elastic scattering process with nuclear targets. Further, since the typical velocity of DM in the Solar neighborhood is of order $\num{e-3}c$, the maximum kinetic energy that can be transferred into the experiment is of order $\num{e-6}m_{\dm}$. For sub-MeV DM, this requires experimental thresholds below the eV scale, meaning that processes such as ionization are not viable channels for DM detection. Moreover, in this regime,
the condensed matter physics of the target becomes important to the DM interaction rate and must be taken into account (for a recent review, see \refcite{Kahn:2021ttr}.)

These challenges have motivated a new generation of direct detection experiments using quantum sensor technologies to achieve extremely low thresholds. In turn, these experiments have spawned a growing literature on the response of target systems to DM interactions~\cite{Hochberg:2021pkt,Boyd:2022tcn,Knapen:2021run,Essig:2015cda,Trickle:2019nya}. These two components have already been combined in several proof-of-principle experiments based on superconducting sensors, notably kinetic inductance detectors (KIDs)~\cite{Gao:2024irf}, transition-edge sensors (TESs)~\cite{Schwemmbauer:2024rcr}, and superconducting nanowire single-photon detectors (SNSPDs)~\cite{Hochberg:2019cyy,Hochberg:2021yud}. These experiments have demonstrated that low background rates can be achieved with energy thresholds low enough to probe new DM parameter space. The future of light DM detection requires us to extend these capabilities to larger exposures and lower energy thresholds.
\begin{figure*}[th!]
    \includegraphics[width=0.98\columnwidth]{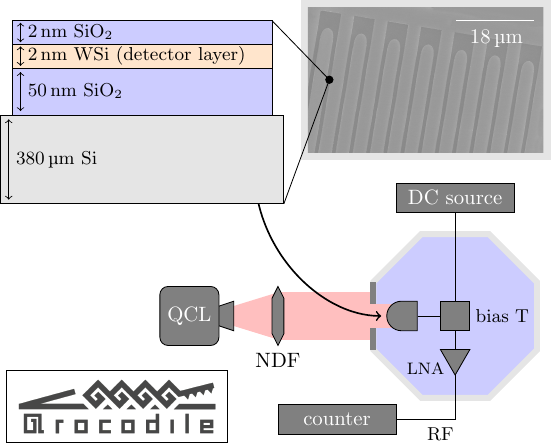}
    \hfill
    \includegraphics[width=0.98\columnwidth]{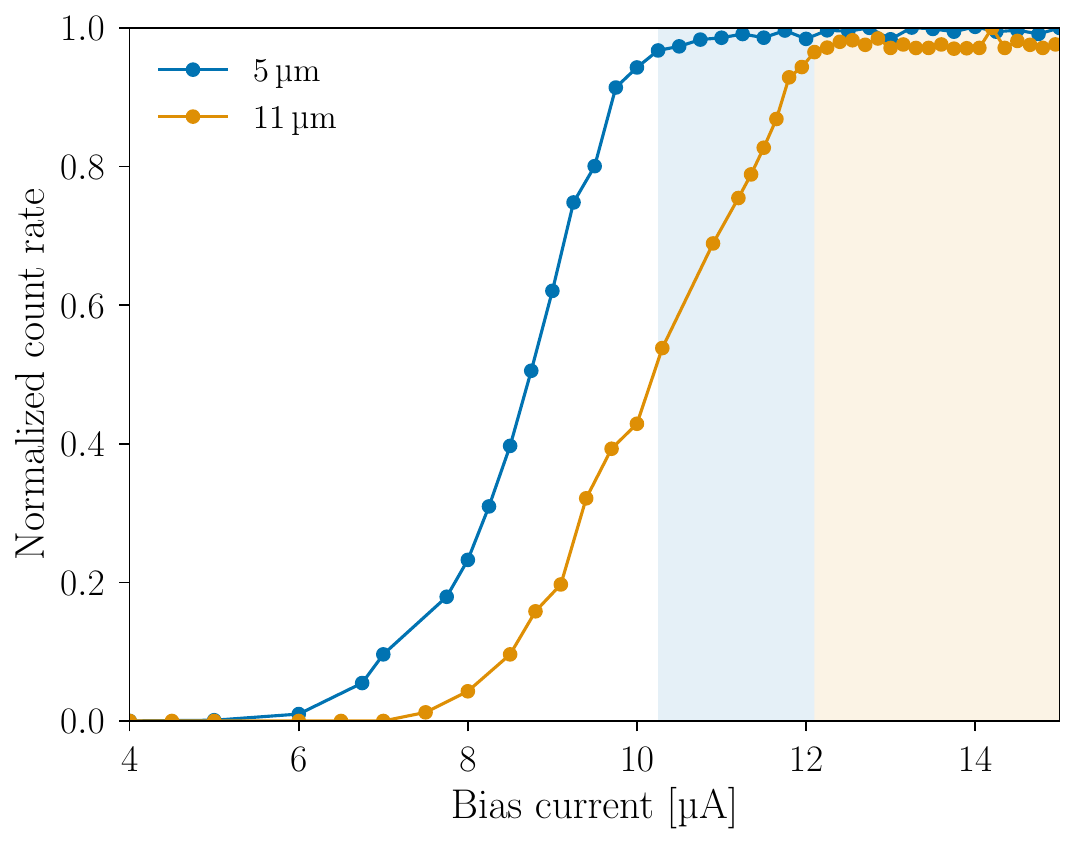}
    \caption{\textbf{The QROCODILE experiment.}
    \textit{Left:} Experimental setup: The detector is mounted into an oxygen-free copper sample holder aligned with the light beam of the quantum-cascade laser (QCL). The initial power of the QCL is significantly reduced by the neutral density filter (NDF). The detector is biased via a high-precision and low-noise DC source. A bias T splits the DC and RF signals from the detector. The RF signal is then amplified with a low-noise amplifier (LNA) to readout via the pulse counter.
    \textit{Inset:} Schematic description of the detector stack. The SEM image represents the meander-shaped SNSPD, with a scale bar of $\qty{18}{\micro\meter}$.
    \textit{Right:} Normalized count rate as a function of the bias current applied to our SNSPD detector under irradiation with \qty{5}{\micro\meter} and \qty{11}{\micro\meter} wavelength photons.}
     \label{fig:setup}
\end{figure*}

In this work, we take a significant step in probing unconstrained DM parameter space with low-threshold quantum sensors. We introduce a new light DM detection experiment, the \textit{Quantum Resolution-Optimized Cryogenic Observatory for Dark matter Incident at Low Energy} (QROCODILE) \footnote{The QROCODILE experiment was formerly known as the \textit{Quantum sensoR cryOgeniC search fOr Dark matter In Light mass rangE} (QROCODILE), but was upgraded in mid-2024 to use optimized capitalization.}. The QROCODILE experiment is based on a microwire-based SNSPD used simultaneously as the target and sensor for DM detection. Our detector has a threshold of \qty{0.11}{\electronvolt}, corresponding to \qty{11}{\micro\meter} photons. We report the first constraints on light DM with this new instrument, placing world-leading limits on sub-MeV DM.

Due to its thin-layer geometry, the QROCODILE sensor is inherently sensitive to the direction of the incoming DM, provided that a sufficient number of signal events are detected. This is crucial for rejecting background and establishing the origin of a putative signal: identifying a candidate signal with the direction of Earth's motion through the Galactic halo has been long recognized as a strong indicator of a DM origin (see e.g. \refcite{Mayet:2016zxu}). Moreover, QROCODILE is also the first experiment to use a superconducting sensor to simultaneously probe the interactions of light DM with both electrons and nucleons. This experiment not only sets leading bounds on DM interactions, but also paves the way for the next generation of quantum sensor--based experiments for DM searches.

\section{QROCODILE anatomy}
\label{sec:experiment}
The QROCODILE experiment consists of a microwire \ce{WSi} SNSPD placed on a \ce{Si}/\ce{SiO2} substrate. The SNSPD functions as both the sensor and the primary target for DM interactions. In principle, the detector is also sensitive to DM interactions in the substrate, but we conservatively do not include these interactions in our analysis. We now detail the design of the experiment, the calibration process, and our science run.

\subsection{Experimental configuration}
\label{sec:setup}
SNSPDs exhibit dramatic sensitivity to mid-infrared photons, with wavelengths as long as \qty{43}{\milli\electronvolt}~\cite{2023Optic..10.1672T}, and high detection efficiency of up to 98\%~\cite{reddy2020superconducting}. Currently, a central challenge in the search for light DM at these low energies is the small effective volume of typical detectors. Built out of narrow nanowires with widths of order \qty{50}{\nano\meter}, these detectors have a limited active area ($\mathord{\sim}10\times\qty{10}{\micro\meter^2}$) due to the complex fabrication process that is required. Additional challenges include complexity in readout signals~\cite{patel2024improvements} and nonuniformity of nanowires that results in suppressed critical current.

To overcome the area limitation, we focused on large-area (${\gg}10\times\qty{10}{\micro\meter^2}$) detectors with wires at least \qty{1}{\micro\meter} wide, which have been proposed~\cite{vodolazov2017single} as an alternative to nanowire-based SNSPDs. Although the detection mechanism of such devices is different from that of nanowire devices~\cite{PhysRevApplied.13.024011}, microwire detectors nonetheless exhibit a similar high internal detection efficiency. 
To date, no measurements beyond the \qty{1550}{\nano\meter} wavelength of \refcite{PhysRevApplied.13.024011} have been reported for detectors based on microwires. To improve our sensitivity in the mid-infrared range, the material stoichiometry was tuned with a higher silicon content, and the experimental current was set higher in ratio to the depairing current (i.e., the theoretical maximum) as suggested by theoretical studies~\cite{vodolazov2017single} and successfully demonstrated in nanowires~\cite{verma2021single,colangelo2022large,2023Optic..10.1672T}.

Our experimental setup is depicted in the left panel of \cref{fig:setup}. The detector was embedded in a \ce{O2}-free copper housing placed at the cold stage of a \qty{100}{\milli\kelvin} dilution cryostat. The active area of the detector is $\qty{600}{\micro\meter}\times\qty{600}{\micro\meter}$ with a wire width of \qty{1}{\micro\meter} (uncertainty of 1.5\%) and filling factor of 25\%. The superconducting layer is composed of \ce{WSi} in the stoichiometric ratio \ce{W45Si55}, to an uncertainty of 1\%, with a transition temperature of \qty{1.73}{\kelvin} and mass of \qty{1.67}ng. The \ce{WSi} layer is encapsulated between two layers of \ce{SiO2} on a \ce{Si} substrate chip of size $\qty{10}{\milli\meter}\times\qty{10}{\milli\meter}$. The inset of the left panel of \cref{fig:setup} shows the detector stack that was optically coupled with mid-infrared light sources.

\subsection{Calibration and science run}
\label{sec:sensitivity}
Before the DM science run, we calibrated the internal detection efficiency of our device to mid-infrared radiation using a quantum-cascade laser setup~\cite{Verma:2020gso} (left panel of \cref{fig:setup}) at wavelengths of \qty{5}{\micro\meter} and \qty{11}{\micro\meter}. The right panel of \cref{fig:setup} shows the photon count rate of our detector normalized to the maximum value as a function of the bias current applied to the device. Importantly, we observed saturation of the count rate at high bias currents, which suggests that the internal detection efficiency is nearly 100\% at both wavelengths~\cite{natarajan2012superconducting}. In SNSPDs, internal detection efficiency refers to the probability that an absorbed photon will produce a detectable electrical signal. When the internal detection efficiency in the SNSPDs is 100\%, it means that every photon that gets absorbed by the nanowire creates a resistive hotspot and triggers a detection event. The bias current has a direct impact on the probability, increasing it as it rises. The sign of 100\% efficiency is a saturation of the photon count rate as a function of the bias current, when applying higher current does not lead to an increase in the count rate, as shown in \cref{fig:setup} (right).

For our DM science run, the optical path was blocked by decoupling the source and sealing the holder to minimize the number of photons reaching the detector. The device was biased with a current of \qty{12.2}{\micro\ampere} and exposed for 415.15 hours at a temperature of \qty{100}{\milli\kelvin}. We recorded 15 individual nonperiodic pulses during the science run, corresponding to a count rate of \qty{e-5}{\per\second}. It is not trivial to determine the origin of these events. The pulse shape is independent of the amount of energy absorbed in the device. Despite the blocked optical paths, the silicon substrate prescreening for radiopurity, and the use of low-radioactivity material for the packaging, 
pulses from sources other than DM may be observed. Cosmic rays and the radioactivity of the surroundings, which are difficult to shield above ground, are potential sources. For future runs, we plan to add muon detectors above and below the cryostat, and we also aim for a full Monte Carlo simulation of the muon-induced event rate. In this work, we simply use the total count rate to place novel constraints on the DM parameter space. While we do not claim that the observed counts constitute a DM signal, neither do we assume that the observed counts can be accounted for by backgrounds. Our constraints are sufficiently conservative to allow for the possibility that all counts originate from DM. Further details on the observed counts and background mitigation strategies are given in the Supplementary Material (SM)~\cite{supplemental-material}, which additionally refers to \refscite{Hochberg:2021yud, Lasenby:2021wsc, Hochberg:2021ymx, SiliconMaterials:3inch, Baudis:2011am, Araujo:2022kip, XENON:2024wpa, Garcia:2022jdt, XENON:2021mrg}.

\section{Dark matter interaction rate}
\label{sec:rates}
The QROCODILE experiment is sensitive to DM interactions through several different mechanisms. We consider: \textbf{(1)} DM-electron scattering, \textbf{(2)} DM absorption onto electrons, and \textbf{(3)} DM-nucleon scattering. In each case, the interaction can take place in the SNSPD sensor itself or in the surrounding substrate. Such an interaction depositing energy above the threshold of the device would cause the SNSPD to register a count. Thus, by evaluating the rates of these processes as a function of DM parameters, we can use the observed count rate to constrain the DM parameter space.

\begin{figure*}\centering
    \includegraphics[width=\textwidth]{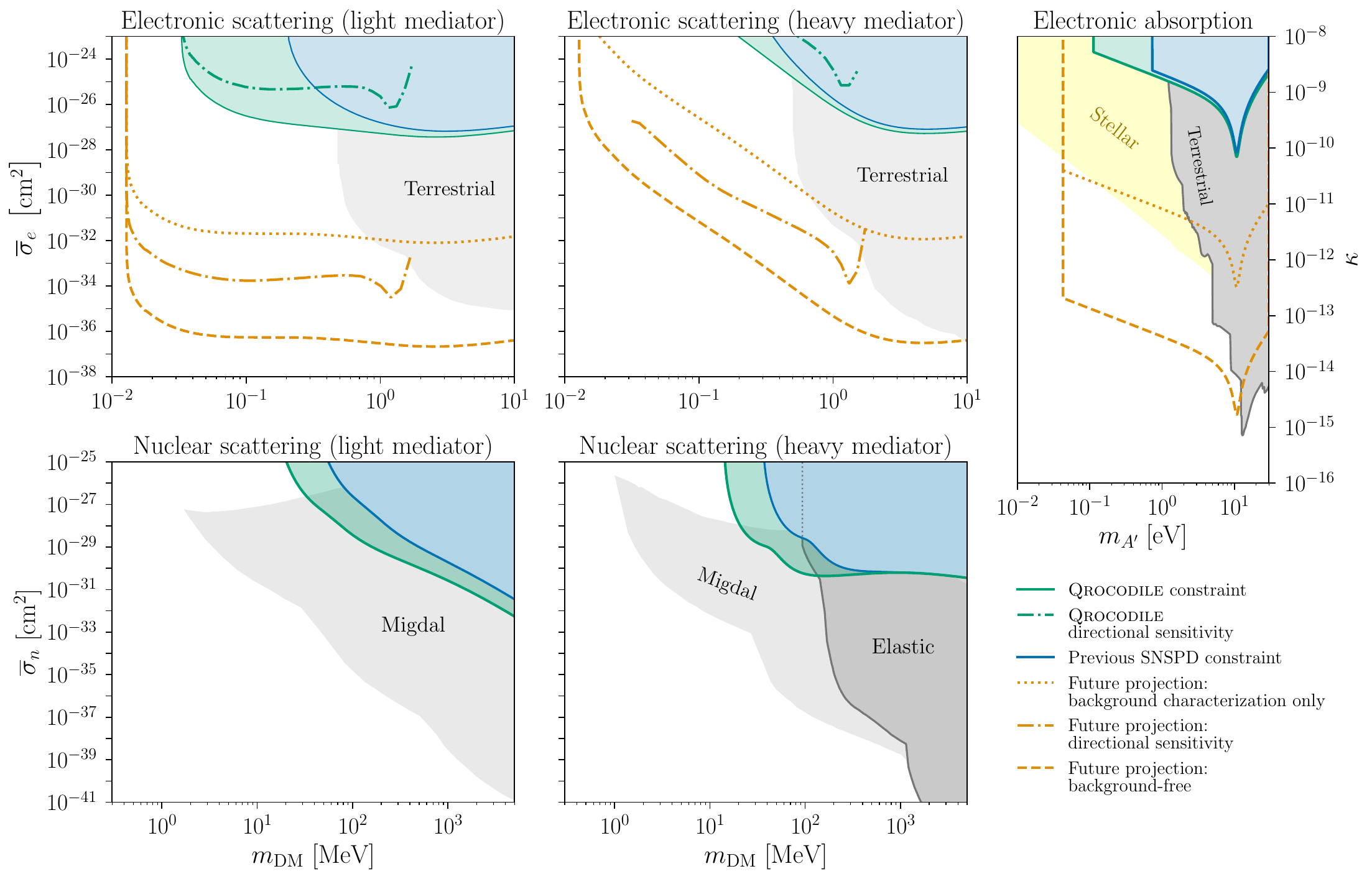}
    \caption{\textbf{Dark matter results.}
        \textit{Top:} new constraints on DM scattering with electrons via a light (\textit{left}) or heavy (\textit{center}) mediator.
        \textit{Bottom:} new constraints on DM scattering with nucleons via a light (\textit{left}) or heavy (\textit{center}) mediator. 
        \textit{Right:} constraints on absorption on electrons of  kinetically-mixed dark photon DM.
        In all panels, green shaded regions indicate the new limits we place using our QROCODILE SNSPD. Blue shaded regions indicate previous SNSPD limits~\cite{Hochberg:2021yud,Hadas:2024}. Dotted orange curves indicate the projected reach of a 10 megapixel SNSPD array with an exposure of one year and a threshold of \qty{29}{\micro\meter} in wavelength, corresponding to \qty{43}{\milli\electronvolt}. Dot-dashed curves indicate cross sections in which the experiment is sensitive to the direction of a DM stream, as a proxy for directional sensitivity, assuming that all events originate from DM. Other existing terrestrial limits from
        \refscite{
        EDELWEISS:2019vjv, EDELWEISS:2022ktt, DarkSide-50:2022qzh, DarkSide_2023, SuperCDMS:2020aus, SuperCDMS:2023sql, CRESST_III_2019, SENSEI:2023zdf, PandaX_4T_2023, LUX:2018akb, XENON_2020,
        Barak:2020fql, Amaral:2020ryn, Aguilar-Arevalo:2019wdi, Essig:2017kqs, Agnes:2018oej, XENON:2019gfn,
        An:2014twa, Agnese:2018col, Aguilar-Arevalo:2019wdi, Arnaud:2020svb, FUNKExperiment:2020ofv, Barak:2020fql}
        are shown in shaded gray, with complementary stellar constraints on absorption~\cite{An:2013yua, An:2014twa, An:2020bxd} appearing in shaded yellow.
        Upper limits of the scattering panels are determined by estimated atmospheric overburden \cite{Emken:2019tni}.
    }
    \label{fig:reach}
\end{figure*}

We compute the rates of these processes following \refscite{Hochberg:2021pkt,Hochberg:2021yud,Hadas:2024}, assuming that the $\dm$ interacts with Standard Model species via a mediator $\med$. The event rate per unit detector mass takes the form
\begin{multline}
    \Gamma =
        \frac{\pi n_\dm\bar\sigma_t}{\mu_{t,\dm}^2}
        \int\frac{\du^3\bb v_\dm\dd^3\bb q\dd\omega}{(2\pi)^3}
        \,f_\dm(\bb v_\dm)
        \\
        \times\mathcal F(q)^2\, S(\bb q, \omega) \delta(\omega - \omega_{\bb q})
    ,
\end{multline}
where $n_\dm$ is the DM number density; $\bar\sigma_t$ is a reference cross section for DM-$t$ scattering, where $t$ denotes the target particle; $\mu_{t,\dm}$ is the reduced mass; $\bb q$ is the 3-momentum transfer; $\omega$ is the energy deposited; $\bb v_\dm$ is the DM speed; $\omega_{\bb q} = \bb q\cdot\bb v_\dm - \bb q^2/2m_\dm$ is the energy transferred at fixed $\bb q$ and $\bb v_\dm$; $f_\dm(\bb v_\dm)$ is the DM velocity distribution function; $\mathcal F(q)$ is a form factor containing the momentum dependence of the interaction potential; and $S(\bb q, \omega)$ is the dynamic structure factor. We assume the standard halo model for $f_\dm(\bb v_\dm)$ with local DM density \qty{0.4}{\giga\electronvolt/\centi\meter^3}, dispersion \qty{230}{\kilo\meter/\second}, escape velocity \qty{600}{\kilo\meter/\second}, and Earth velocity \qty{240}{\kilo\meter/\second} in the Galactic frame~\cite{Lewin:1995rx,Stanic:2025yze}. The form factor is given by $\mathcal F(q) = (m_\phi^2+q_{0,t}^2)/(m_\phi^2+q^2)$ for scattering via a mediator of mass $m_\phi$ with a reference momentum $q_{0,t}$. We take $q_{0,e} \equiv \alpha m_e$ and $q_{0,n} \equiv m_\dm\langle v_\dm\rangle$ for electronic and nuclear scattering, respectively. In both cases, we take $\bar\sigma_t \equiv \frac1\pi\mu_{t,\dm}^2g_0^2/(m_\med^2 + q_{0,t}^2)^2$, defining $g_0$ to absorb couplings.
Finally, the dynamic structure factor is determined by the available final states $\left|f\right\rangle$ of the target system, and is given by
\begin{equation}
    \label{eq:dynamic-structure-factor}
    S(\bb q,\omega) \equiv \frac{2\pi}{V}\sum_{f}
    \bigl|\left\langle f\middle|\,\hat n(-\bb q)\,\middle|0\right\rangle\bigr|^2
    \delta\bigl(\omega - \left[E_f-E_0\right]\bigr)
    .
\end{equation}
Computing event rates now reduces to computing the dynamic structure factor for each channel.

For DM interactions with electrons (i.e., cases 1 and 2 above), we use the linear response theory of dielectric systems, as detailed in \refcite{Hochberg:2021pkt}. For spin-independent scattering, the dynamic structure factor is given by the fluctuation-dissipation theorem as $S(\bb q, \omega) = 2 \Im\chi(\bb q, \omega)$, where $\chi$ is the linear response function of the electron number density. The linear response function also determines the absorption rate for the case of kinetically-mixed dark photon DM (see e.g. \refcite{Fabbrichesi:2020wbt}), where the interaction Lagrangian has the form $\mathcal L_{\mathrm{int}} = -\frac12\kappa F_{\mu\nu}F^{\prime\mu\nu}$ with $F_{\mu\nu}^{(\prime)} \equiv \partial_\mu A^{(\prime)}_\nu - \partial_\nu A^{(\prime)}_\mu$, where $A_\mu$ is the photon field and $A'_\mu$ is dark photon field. Given this particular interaction structure, the absorption rate is $\Gamma_{\mathrm{A}} = m_\dm\kappa^2e^2\bb p_\dm^{-2}\Im\chi(\bb p_\dm, m_\dm)$, where $\bb p_\dm = m_\dm\bb v_\dm$~\cite{Hochberg:2021yud}.
The response function $\chi$ can be readily computed analytically using the random phase approximation (RPA) in the zero-temperature limit, which is appropriate for our case: the temperature of \qty{100}{\milli\kelvin} in the experiment is well below all energy scales of the material. In the future, systematic uncertainty in the modeling of the response function could be reduced by experimental calibration or by dedicated density functional theory computations.

Crucially, the linear response function can be modified by the geometry of the system. In particular, the standard RPA response function assumes that the interaction takes place in an infinite bulk volume. This approximation fails once the shortest length in the target system becomes smaller than the inverse momentum transfer in the process. SNSPDs are thin-layer devices, so for small DM masses, with correspondingly small momenta, the response function receives geometric corrections that can substantially enhance the DM interaction rate~\cite{Lasenby:2021wsc,Hochberg:2021yud}.

The previous generation of low-threshold DM scattering searches has been sensitive to DM masses above \qty{200}{\kilo\electronvolt}~\cite{Hochberg:2021yud}, with typical inverse momenta of order \qty{1}{\nano\meter} or shorter, still much smaller than the layer size of $\mathcal O(\qty{10}{\nano\meter})$. Here geometric considerations were irrelevant. However, QROCODILE is sensitive to DM masses as low as \qty{30}{\kilo\electronvolt}, with $1/q\sim\qty{10}{\nano\meter}$, and the depth of the detector layer has been reduced to \qty{2}{\nano\meter}. Thus, geometric effects are non-negligible at low masses.

In the parameter space probed by QROCODILE, the geometric enhancement to the overall scattering rate is expected to be $\mathcal O(1)$, and we conservatively do not include it in our constraints. Crucially, however, the enhancement is anisotropic in the momentum transfer, and thus gives rise to directional sensitivity. This is an essential tool for rejecting backgrounds and confirming the origin of any candidate DM signal~\cite{Mayet:2016zxu,Hochberg:2021ymx,Boyd:2022tcn}: since Earth has a direction of motion through the galactic DM halo, there is a known preferred direction for the motion of DM particles in the lab frame, known as the DM `wind.' It is extremely difficult for backgrounds to supply the same directional dependence. We therefore compute the anisotropy in the scattering rate by numerically solving Maxwell's equations with the appropriate boundary conditions, as detailed in \refscite{Hochberg:2021yud,Lasenby:2021wsc}. We use this anisotropy to identify cross sections at which QROCODILE is sensitive to the DM wind, using cold streams in different directions as a proxy for the modulation in the rate as Earth rotates over the course of a sidereal day. Further details on both bulk and thin-layer responses are given in the SM.

We can also consider the scattering of DM particles with nuclei. While the SNSPD sensor is nominally sensitive to the dissociation of Cooper pairs---a process that takes place in the electron system---the sensor can nonetheless be triggered by a nuclear scattering event via phonon production, as detailed by \refcite{Hadas:2024}. In the elastic scattering regime, the energy that can be transferred from the DM to a nucleon of mass $m_N \gg m_\dm$ is sharply limited. Historically, this played a role in motivating electron recoil experiments, with a target mass $m_\el \ll m_N$. However, at small energy transfers, the process is sensitive to the structure of the lattice, and the kinematics of phonons and quasiparticles can be much more favorable to DM scattering. 
Here we place a conservative limit on DM interactions with nuclei via nuclear recoils, where the dynamic structure factor is given by~\cite{Trickle:2019nya}:
\begin{equation}
    \label{eq:structure-factor-elastic}
    S(\bb q, \omega) = \frac{2\pi\rho_{\mathrm{T}}}{\sum_{N} A_{N}} \sum_{N} \frac{A_N^3}{m_N}
    F_N(\bb q)
    \delta \left(\omega - \frac{\bb q^2}{2 m_N}\right)
    \,.
\end{equation}
Here $N$ indexes the nuclei in a unit cell; $m_N$ is the atomic mass; $A_N = m_N/\qty{}{u}$ is the atomic mass number; $f_n$ is the coupling to DM; and $F_N(\bb q)$ is the nuclear form factor. For the latter we take the Helm form factor~\cite{Helm:1956zz}, $F_N (q) = [3 j_1 (q r_N)/(q r_N)] e^{-(qs)^2/2}$,  with $q=|{\bb q}|$, $j_1$ the spherical Bessel function of the first kind, $r_N \approx A_N^{1/3}\times\qty{1.14}{\femto\meter}$  the effective nuclear radius, and $s$ the nuclear skin thickness. We use $\{A_{\ce{W}}, A_{\ce{Si}}\} \approx \{183.85, 28.09\}$, and we take $s = \qty{0.9}{\femto\meter}$~\cite{Helm:1956zz,Trickle:2019nya}.
The low threshold of our device allows, in principle, for limits to be placed at even lower DM masses through the production of multi-phonons. The reach is then determined by the vibrational spectrum for the amorphous \ce{WSi} used in our detector layer, and will appear in a separate publication~\cite{QROCODILE_future}.

\section{Constraints and discussion}
\label{sec:results}
The constraints placed by QROCODILE on the DM parameter space are shown in \cref{fig:reach}. The top-left and top-center panels show the results for DM-electron scattering in the limit of a light or heavy mediator, respectively. The bottom-left and bottom-center panels shows results for DM-nucleon scattering via a light and heavy mediator, respectively, utilizing the nuclear-recoil channel only. The right panel shows constraints on electronic absorption of dark photon DM. Our new constraints are indicated by shaded green, while previous SNSPD limits~\cite{Hochberg:2021yud} are shown in shaded blue. Our bounds are set at the 95\% confidence level, incorporating the measured count rate via the Feldman-Cousins procedure~\cite{Feldman:1997qc}\footnote{This is the same limit that would be obtained if one assumed that all observed counts originate from DM. Assuming that they originate from backgrounds would result in a less conservative limit.}. Shaded gray regions indicate existing terrestrial constraints from \refscite{
        EDELWEISS:2019vjv, EDELWEISS:2022ktt, DarkSide-50:2022qzh, DarkSide_2023, SuperCDMS:2020aus, SuperCDMS:2023sql, CRESST_III_2019, SENSEI:2023zdf, PandaX_4T_2023, LUX:2018akb, XENON_2020,
        Barak:2020fql, Amaral:2020ryn, Aguilar-Arevalo:2019wdi, Essig:2017kqs, Agnes:2018oej, XENON:2019gfn,
        An:2014twa, Agnese:2018col, Aguilar-Arevalo:2019wdi, Arnaud:2020svb, FUNKExperiment:2020ofv, Barak:2020fql}, and the shaded yellow region indicates model-dependent constraints from stellar cooling~\cite{An:2013yua, An:2014twa, An:2020bxd}.
We set the upper limits of the scattering panels to reflect estimated cross sections at which atmospheric overburden becomes relevant (see e.g.~\cite{Emken:2019tni}).

For DM-electron scattering, the dot-dashed green curve shows the region in which the detector would be sensitive to the direction a DM stream, which is a proxy for the directional sensitivity of the device. The directional sensitivity arises from the anisotropy in the geometry of the device itself, which modifies the dielectric response along different directions (see SM for details). Since the number of total events in our dataset is negligible, we are able to set a more restrictive bound on the basis of the count rate alone. With only 15 events, a very large anisotropy in the rate would be needed to have a high probability of detecting a modulation. Thus, as expected, the timestamps of our events are consistent with an isotropic background at the 95\% confidence level. However, in future experimental runs with larger exposures, background counts will limit the scaling of this constraint, whereas the directional sensitivity scales more favorably.

Our new results surpass previous upper limits on the cross-section, setting stronger constraints in an unexplored DM mass range.
Our constraints on nuclear scattering lie in parameter space that is also probed by the Migdal effect in semiconductors, but our constraints on DM-electron scattering are the first bounds in this portion of parameter space independent of the Migdal effect. The relative sensitivity of the QROCODILE experiment is particularly pronounced for DM-electron scattering via a light mediator. Here, our results provide the first nontrivial constraints on DM interactions at masses as low as \qty{30}{\kilo\electronvolt}.

For our DM-electron scattering and absorption results, we also show the projected reach of a future experiment with a similar configuration consisting of \num{e7} subunits (i.e., pixels), each with the same size and composition as our prototype device. For this future experiment, we assume a threshold sensitivity of \qty{29}{\micro\meter} (\qty{43}{\milli\electronvolt}). An SNSPD with this threshold has already been demonstrated by \refcite{2023Optic..10.1672T}, and will allow us to probe significantly lower DM masses and cross sections. While it is difficult to estimate the irreducible dark count rate in a scaled detector, we show two dashed lines corresponding to no background (`background-free') and the scenario in which the number of background counts scales linearly with the exposure from the number observed in the current science run, assuming only statistical uncertainty in the background rate (`background characterization only'). We also indicate directional sensitivity for this configuration. Here, the directional sensitivity exceeds the background-limited sensitivity with raw counts, providing orders of magnitude of additional reach. The directional detection sensitivity of SNSPDs thus places our experiment in a unique position: 
QROCODILE is capable not only of excluding parameter space, but of discriminating betweeen backgrounds and a modulating signal, thus allowing to establish a DM discovery.

While our SNSPD detector has demonstrated high internal detection efficiency with low noise, there are several strategies to further improve sensitivity. Firstly, the experiment can be carried out underground to adequately protect the setup from cosmic rays. With a longer exposure, the existing limits should be significantly improved. Secondly, the energy threshold can be further reduced to the already demonstrated level of \qty{43}{\milli\electronvolt}, which is still far from the fundamental limit closer to the superconducting gap of $\mathcal O(\qty{}{\milli\electronvolt})$. This will require further optimization of the stoichiometry~\cite{verma2021single}. The detector mass can be increased by increasing the sensor area and the wire width. The first is limited by the kinetic inductance~\cite{kerman2006kinetic} of the superconducting materials while the second is restricted by the Pearl length~\cite{vodolazov2017single}. The QROCODILE Collaboration plans to pursue work along these lines in order to take an even bigger bite into DM parameter space with our next science run, the upcoming \textit{Next Incremental Low-threshold Exposure} (NILE QROCODILE).

\textbf{Acknowledgments.} We thank Ben Kilminster for useful discussions, Rotem Ovadia for helpful input, and Francesco Ferella from the LNGS Chemistry Service for performing the HR-ICP-MS of the \ce{Si} substrate raw material. We are grateful to Maxim Karmantzov for our logo design. 
The work of Y.H. is supported by the Israel Science Foundation (grant No. 1818/22), by the Binational Science Foundation (grants No. 2018140 and No. 2022287) and by an ERC STG grant (``Light-Dark,'' grant No. 101040019). 
The work of B.V.L. is supported by the MIT Pappalardo Fellowship.
This project has received funding from the European Research Council (ERC) under the European Union’s Horizon Europe research and innovation programme (grant agreement No. 101040019) and from the University of Zurich.  Views and opinions expressed are however those of the author(s) only and do not necessarily reflect those of the European Union. The European Union cannot be held responsible for them. 

\bibliography{references}


\onecolumngrid{}
\clearpage

\setcounter{page}{1}
\setcounter{equation}{0}
\setcounter{figure}{0}
\setcounter{table}{0}
\setcounter{section}{0}
\setcounter{subsection}{0}
\renewcommand{\theequation}{S.\arabic{equation}}
\renewcommand{\thefigure}{S\arabic{figure}}
\renewcommand{\thetable}{S\arabic{table}}
\renewcommand{\thesection}{\Roman{section}}
\renewcommand{\thesubsection}{\Alph{subsection}}
\newcommand{\ssection}[1]{
    \addtocounter{section}{1}
    \oldsection{\thesection.~~~#1}
    \addtocounter{section}{-1}
    \refstepcounter{section}
    \noindent\ignorespaces{}
}
\newcommand{\ssubsection}[1]{
    \addtocounter{subsection}{1}
    \subsection{\thesubsection.~~~#1}
    \addtocounter{subsection}{-1}
    \refstepcounter{subsection}
    \noindent\ignorespaces{}
}
\newcommand{\fakeaffil}[2]{$^{#1}$\textit{#2}\\}

\thispagestyle{empty}
\begin{center}
    \begin{spacing}{1.2}
        \textbf{\large 
            Supplemental Material:\\
            First Sub-MeV Dark Matter Search with the QROCODILE Experiment Using Superconducting Nanowire Single-Photon Detectors
        }
    \end{spacing}
    \par\smallskip
    Laura Baudis,\textsuperscript{1}
    Alexander Bismark,\textsuperscript{1}
    Noah Brugger,\textsuperscript{1}
    Chiara Capelli,\textsuperscript{1}
    Ilya Charaev,\textsuperscript{1}\\
    Jose Cuenca Garc\'\i{}a,\textsuperscript{1}
    Guy Daniel Hadas,\textsuperscript{2}
    Yonit Hochberg,\textsuperscript{2,\,3}
    Judith K. Hohmann,\textsuperscript{4,\,5}\\
    Alexander Kavner,\textsuperscript{1}
    Christian Koos,\textsuperscript{4}
    Artem Kuzmin,\textsuperscript{4}
    Benjamin V. Lehmann,\textsuperscript{5}
    Severin \\N\"ageli,\textsuperscript{1}
    Titus Neupert,\textsuperscript{1}
    Bjoern Penning,\textsuperscript{1}
    Diego Ram\'\i{}rez Garc\'\i{}a,\textsuperscript{1}
    and Andreas Schilling\textsuperscript{1}
    \par\smallskip
    {\small
        \fakeaffil{1}{Department of Physics, University of Z\"urich, Winterthurerstrasse 190, CH-8057 Z\"urich, Switzerland}
        \fakeaffil{2}{Racah Institute of Physics, Hebrew University of Jerusalem, Jerusalem 91904, Israel}
        \fakeaffil{3}{Laboratory for Elementary Particle Physics, Cornell University, Ithaca, NY 14853, USA}
        \fakeaffil{4}{Karlsruhe Nano Micro Facility (KNMFi), Engesserstrasse 5, 76131 Karlsruhe, Germany}
        \fakeaffil{5}{Karlsruhe Institute of Technology, Engesserstrasse 5, 76131 Karlsruhe, Germany}
        \fakeaffil{6}{Center for Theoretical Physics, Massachusetts Institute of Technology, Cambridge, MA 02139, USA}
        (Dated: \today)
    }

\end{center}
\par\smallskip

In this Supplemental Material, we provide further details on the dark matter (DM) scattering rate computations, including directional sensitivity from the geometry of the device, and on the fabrication and data analysis of the detector. The numbering of references matches that of the main text.

\ssection{Dark matter scattering rate}
To compute the DM scattering rate with electrons for spin-independent interactions in the bulk limit, we take $S(\bb q, \omega) = 2 \Im\chi(\bb q, \omega)$, where $\chi$ is the linear response function of the electron density. We compute $\chi$ in terms of the dielectric function $\epsilon$, since
\begin{equation}
    \chi(\bb q, \omega) = \frac{1}{V_{\mathrm{C}}(q)}\frac{1}{\epsilon(\bb q, \omega)}
    ,
\end{equation}
where $V_{\mathrm{C}}(q) = e^2/q^2$ is the Coulomb potential. We approximate $\epsilon(\bb q, \omega)$ by the Lindhard dielectric function $\epsilon_\lind(\bb q, \omega)$. We consider only deposits in the \ce{WSi} detector layer, for which we assume a Fermi energy of $E_\fermi=\qty{7}{\electronvolt}$. The plasma frequency is taken to be $\omega_p = \lambda_{\mathrm{TF}}^{-1}v_{\fermi}/\sqrt{3} \approx \qty{10.8}{\electronvolt}$, where the inverse Thomas-Fermi length is given by $\lambda_{\mathrm{TF}}^{-1} = (e/\pi)\sqrt{m_\el k_\fermi}$. We take the plasmon to have a width of $\Gamma = 0.1\times E_\fermi$.

\begin{figure}
    \includegraphics[width=0.32\columnwidth]{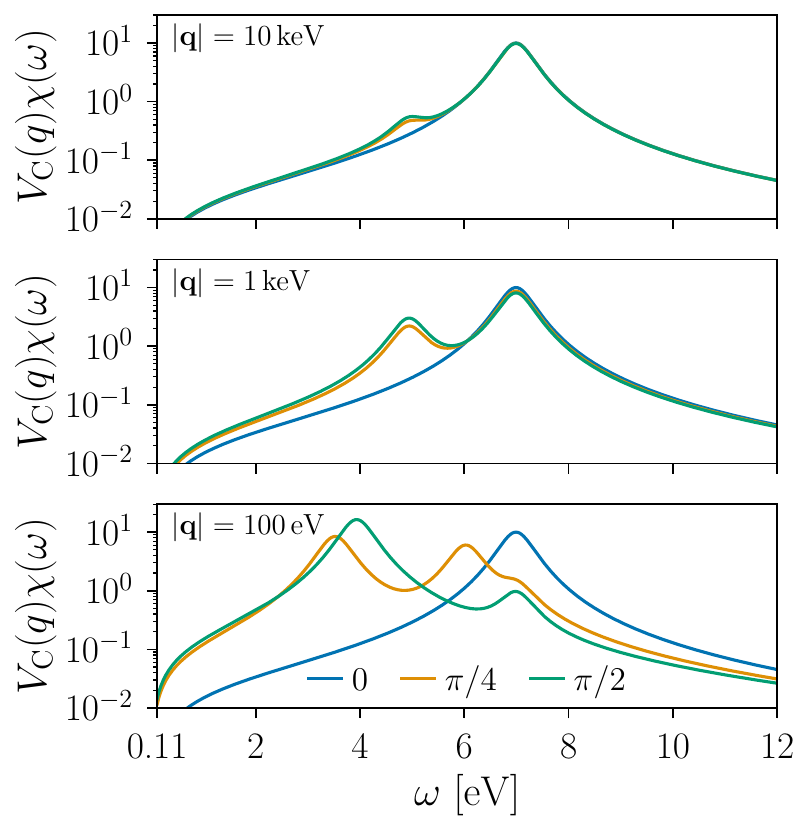}
    \hfill
    \includegraphics[width=0.32\columnwidth]{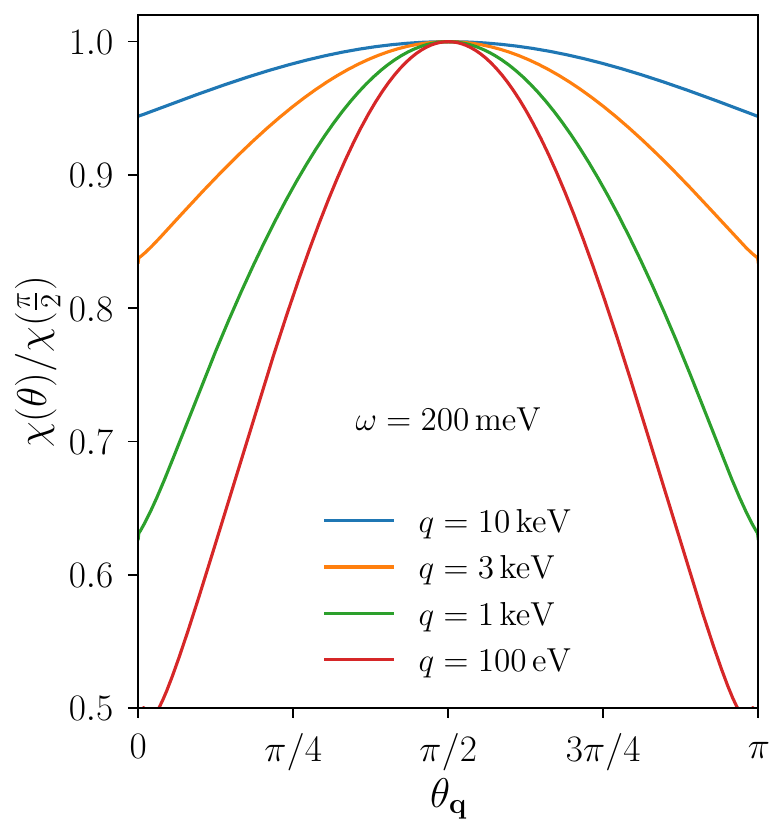}
    \hfill
    \includegraphics[width=0.32\columnwidth]{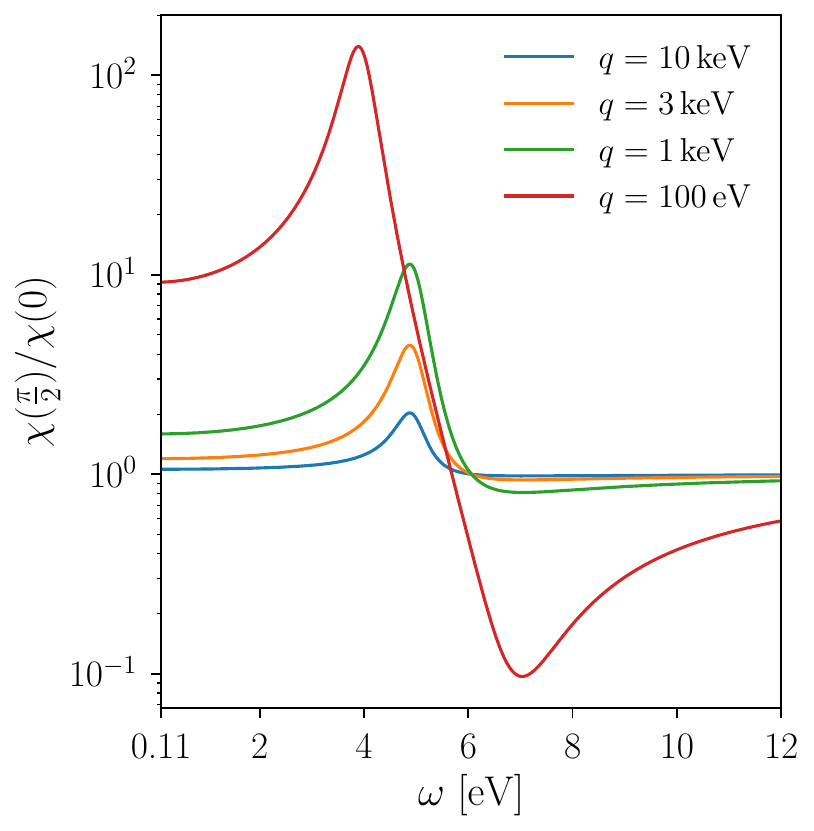}\\
    \includegraphics[width=\columnwidth]{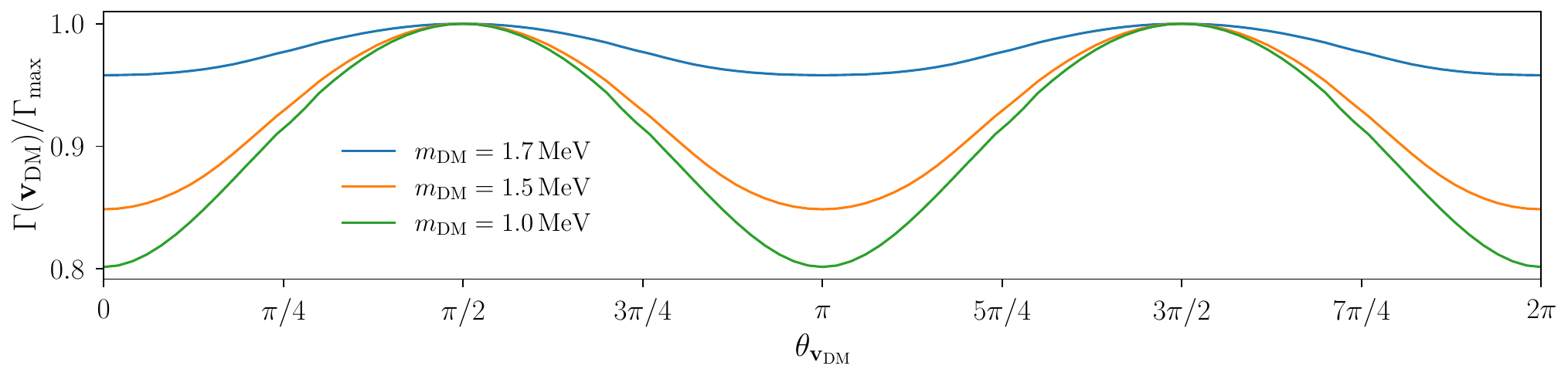}
    \caption{\textbf{Anisotropic response function.}
    \textit{Top left:} the linear response function $\chi(\bb q, \omega)$ as a function of energy for three different magnitudes and directions of momentum transfer. The angle $\theta_{\bb q}$ is measured from the normal direction of the detector layer to the direction of $\bb q$.
    \textit{Top center:} angular dependence of $\chi$ at fixed energy and momentum transfer. The response function is plotted in ratio to its value at $\theta_{\bb q}=\pi/2$ for ease of comparison across different curves.
    \textit{Top right:} the ratio of the response function at its maximum ($\theta_{\bb q}=\pi/2$) to its minimum ($\theta_{\bb q}=0$) at fixed momentum transfer, as a function of energy.
    \textit{Bottom:} modulation in the scattering rate for DM at fixed speed as a function of the angle $\theta_{\bb{v}_{\dm}}$ between the DM velocity and the normal direction of the detector layer.}
    \label{fig:anisotropic-response}
\end{figure}

In the case of thin layers, we apply the procedure from \refscitesm{Hochberg:2021yud,Lasenby:2021wsc}, and refer the reader to these works for details. Briefly, the response function is extracted from the solution to Maxwell's equations for the electric potential $\varphi(\bb x, t)$ in the presence of a periodic source of the form $\rho = \rho_0e^{i\bb q\cdot\bb x-i\omega t}$. We take the layer to lie in the $xy$ plane, and define $\psi(z)$ via the ansatz $\varphi(\bb x, t) = \psi(z) e^{i\bb q\cdot\bb x-i\omega t}$. The dynamic structure factor then takes the form
\begin{equation}
    \label{eq:thin-loss}
    S(\bb q, \omega) = \frac12\frac{q^2}{d}\Re\left[
        -i\frac{1}{\rho_0}\int\du z\left(
            i\psi(z) + \frac{q_z}{q^2}\psi'(z)
        \right)
    \right]
    .
\end{equation}
We solve for $\psi(z)$ by substituting directly into Maxwell's equations, taking $\epsilon(z)$ to be a piecewise function different in each layer. We model the detector as a single layer of \ce{WSi} in vacuum, and we model the dielectric function of \ce{WSi} with the Drude function, as is appropriate for small momentum transfers. The resulting response function is dependent on the direction of the momentum transfer $\bb q$, as shown in \cref{fig:anisotropic-response}. Here, the anisotropy becomes very large at momentum transfers $q \sim \mathcal O(1/\qty{}{\nano\meter})$, or about \qty{100}{\electronvolt}, and already appears significant at $q \sim \qty{1}{\kilo\electronvolt}$ for energies above \qty{4}{\electronvolt}. However, since the typical energy deposit is $\omega \sim qv_\dm \sim \num{e-3}\times q$, anisotropic scattering is relevant only for $\omega \lesssim \qty{1}{\electronvolt}$, corresponding to DM masses $m_\dm \lesssim \qty{1}{\mega\electronvolt}$. The direction of $\bb q$ is correlated with the direction of $\bb v_\dm$, so sensitivity of the rate to the direction of $\bb q$ yields sensitivity to the direction of $\bb v_\dm$. Since $\bb q$ and $\bb v_\dm$ are only correlated, and not perfectly aligned, the anisotropy of the rate with respect to the DM velocity is somewhat suppressed compared to the anisotropy with respect to $\bb q$, as shown in the bottom panel of \cref{fig:anisotropic-response}.

Direct evaluation of the scattering rate using the full velocity distribution is very computationally expensive. There are three independent directions in the problem: those of the momentum transfer $\bb q$, the velocity $\bb v$, and the normal to the plane of the layer, $\bb{\hat z}$. This eliminates the symmetries that are usually used to simplify the costly integrals in the evaluation of the rate. We thus employ the following approximation to compute the daily modulation. We treat the DM as a perfectly cold stream with fixed velocity, and compute the rate only for $\bb v \parallel \bb{\hat z}$ and $\bb v \perp \bb{\hat z}$. We estimate the modulation as the ratio of these two rates. In practice, due to the dispersion in the DM velocity distribution, the modulation is slightly reduced.

Given the daily modulation, we compute the number of events needed to statistically distinguish a directional signal. This leads to an anisotropic reach, defined following \refcitesm{Hochberg:2021ymx} as the cross section at which a directional signal would be discernible at 95\% CL. In particular, we determine the number of events such that the rates in the directions parallel and perpendicular to the DM wind can be distinguished. We treat the event count in each direction as a Poisson random variable, and use the fact that the difference of the two Poisson variables is Skellam-distributed.

\begin{figure}\centering
    \sbox\twosubbox{%
        \resizebox{\dimexpr.95\textwidth-1em}{!}{%
            \includegraphics[trim={0 0 0 0},clip,height=3cm]{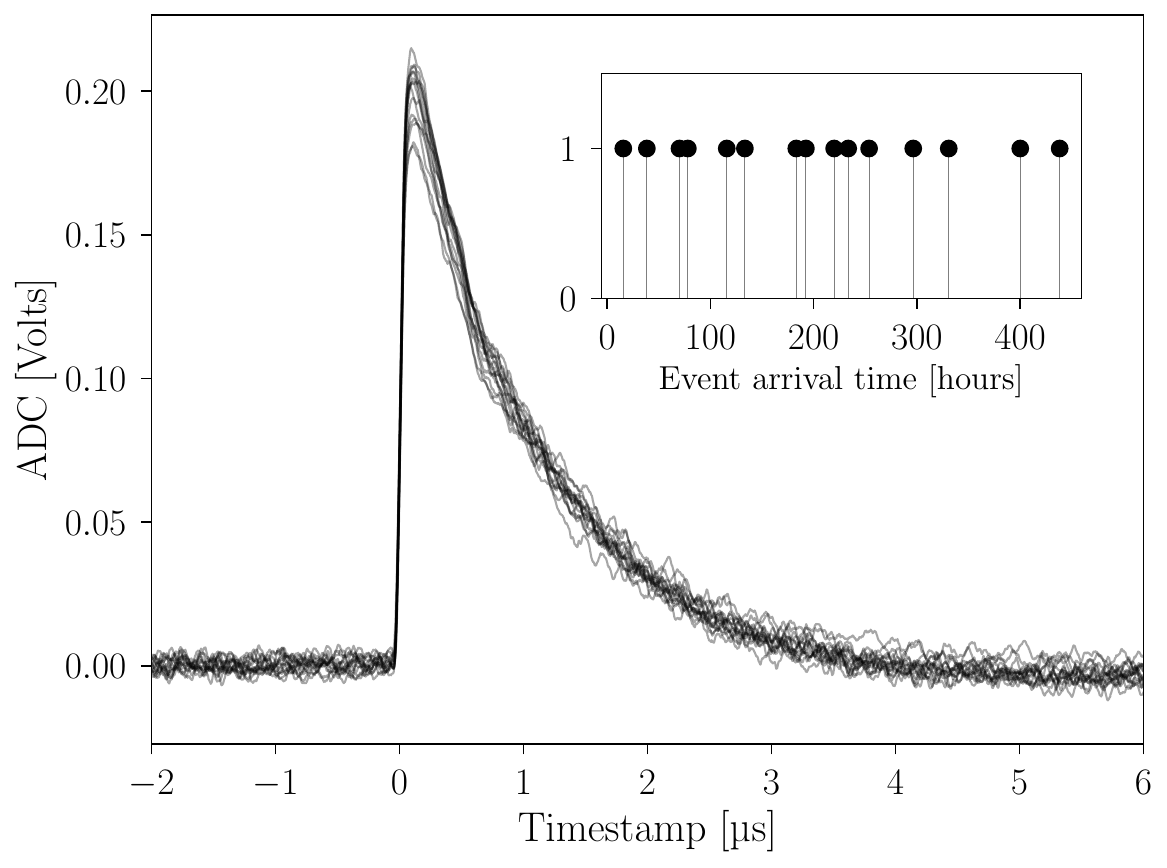}%
            \includegraphics[trim={0 0 0 0},clip,height=3cm]{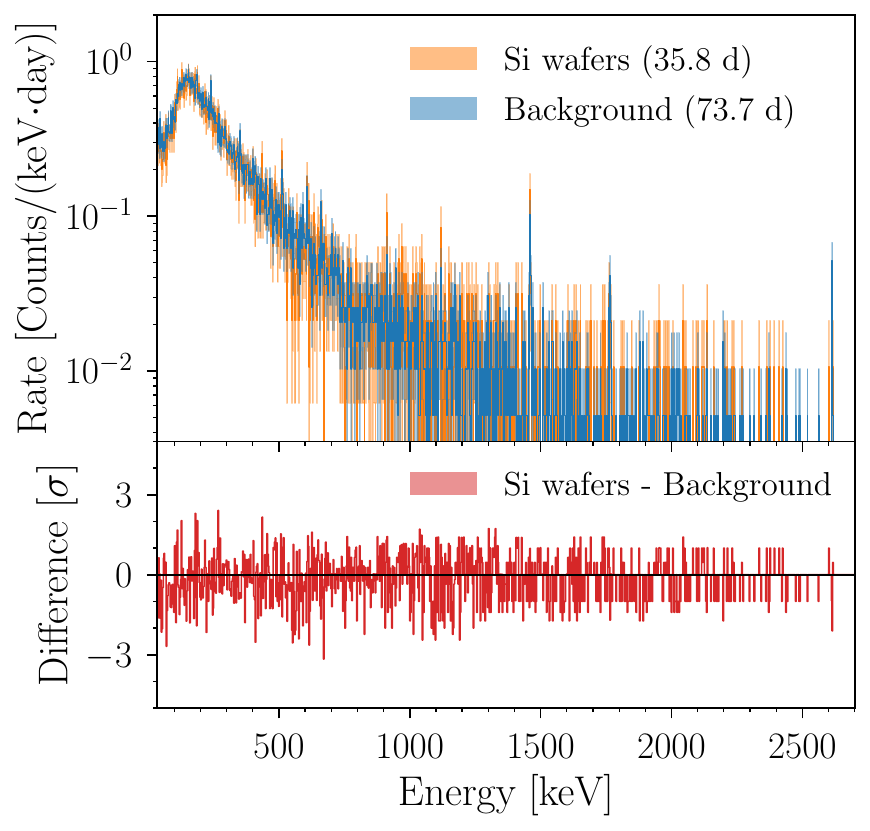}%
        }%
    }
    \setlength{\twosubht}{\ht\twosubbox}
    
    \includegraphics[height=\twosubht]{waveforms.pdf}
    \hfill
    \includegraphics[height=\twosubht]{gator_spectrum_si_wafers_nolines_compact.pdf}    
    \caption{\textit{Left:} \textbf{Waveforms.} The 15 measured waveforms recorded over the DM science run drawn superimposed.
    \textit{Inset:} arrival time of each event throughout the run.
    \textit{Right:} \textbf{Energy spectrum.} Energy spectrum of the silicon substrate wafers (orange), measured with the Gator low-background germanium spectrometer, compared to the background spectrum (blue). Statistical uncertainties are indicated as transparent bands. The background-subtracted spectrum, normalized to the combined uncertainty in units of standard deviations, is given in red. 
    } 
    \label{fig:waveforms}
\end{figure}

\ssection{Detector fabrication and analysis}\label{app2}
The device was fabricated from a \qty{2}{\nano\meter} thick \ce{WSi} (silicon-rich face) film, sputtered onto a \ce{SiO2}/\ce{Si} substrate at room temperature using RF co-sputtering. To prevent oxidation of the superconductor, a \qty{2}{\nano\meter} \ce{Si} layer was deposited \textit{in situ} on top of the \ce{WSi} film. Microwires were patterned using electron beam lithography with a high-resolution positive e-beam resist. The ZEP 520 A resist was spin-coated onto the chip at \qty{5500}{rpm}, yielding a thickness of \qty{320}{\nano\meter}. After exposure, the resist was developed by immersing it in O-xylene for 50 seconds, followed by a rinse in 2-propanol. The ZEP 520 A pattern was then transferred to the \ce{WSi} through reactive ion etching in CF4 at \qty{60}{\watt} for 4 minutes. The fabricated detectors were sorted according to their absolute critical current values. Samples with the highest values at \qty{100}{\milli\kelvin} were optically characterized under mid-infrared irradiation. The detector with saturated internal quantum efficiency was used in the DM science run.  We show the measured waveforms during the DM science run along with their arrival times in the left panel of \cref{fig:waveforms}.

\begin{table}
    \begin{tabular*}{\textwidth}{@{\extracolsep{\fill}} l|cccccccccc}
        \toprule
        & \textbf{\ce{^{238}U}}
        & \textbf{\ce{^{226}Ra}}
        & \textbf{\ce{^{228}Ra}}
        & \textbf{\ce{^{228}Th}}
        & \textbf{\ce{^{235}U}}
        & \textbf{\ce{^{60}Co}}
        & \textbf{\ce{^{40}K}}
        & \textbf{\ce{^{137}Cs}}
        & \textbf{\ce{^{54}Mn}}
        & \textbf{\ce{^{58}Co}} \\
        \midrule
        \ce{Si} substrate
        & $ < 99.1 $
        & $ < 7.85 $
        & $ < 14.5 $
        & $ < 12.1 $
        & $ < 4.40 $
        & $ < 2.13 $
        & $ < 33.6 $
        & $ < 2.32 $
        & $ < 1.62 $
        & $ < 1.60 $ \\
        Copper
        & $ < 1.06 $
        & $ < 0.21 $
        & $ < 0.08 $
        & $ < 0.01 $
        & $ - $
        & $ \num{0.08(1)}$
        & $ < 0.42$
        & $ < 0.011$
        & $ - $
        & $ - $ \\
        \bottomrule
    \end{tabular*}
    \caption{\textbf{Material activity.} Measured specific activities in units of \qty{}{\milli\becquerel/\kilo\gram} with $\pm 1\sigma$ uncertainties and upper limits at 90\% (95\%) C.L. for the \ce{Si} substrate (oxygen-free copper) material as obtained with $\gamma$-spectrometry at the Gator (GeMSE) facility.}
    \label{tab:gator_activities}
\end{table}

To ensure a low background rate from radioactivity, the Si substrate raw material underwent radioassay to quantify the activity of common radioactive impurities. Twelve 3-inch silicon wafers from Silicon Materials Inc.~\citesm{SiliconMaterials:3inch}, amounting to a sample mass of about \qty{48}{\gram}, were screened in the Gator high-purity germanium, low-background counting facility~\citesm{Baudis:2011am, Araujo:2022kip}. The corresponding $\gamma$-ray spectrum from a sample measurement with a live time of  \qty{35.8}{\day}, together with the background spectrum used in the activity analysis, is displayed in the right panel of \cref{fig:waveforms}. Only upper limits on the activity were obtained for all investigated isotopes or decay chains, as summarized in \cref{tab:gator_activities}. To further enhance the abundance estimate for selected impurities, a high-resolution inductively coupled plasma mass spectrometric (HR-ICP-MS) analysis of a subset of the sample measured with Gator was performed by the LNGS Chemistry Service. Concentrations of $\ce{K} < \qty{5}{ppm}$, $\ce{Th} < \qty{5}{ppb}$, and $\ce{U} < \qty{1}{ppb}$ were obtained. This corresponds to specific activities of $\ce{^{40}K} < \qty{160}{mBq/kg}$, $\ce{^{228}Ra} < \qty{20}{mBq/kg}$, and $\ce{^{238}U} < \qty{12}{mBq/kg}$, further validating the radiopurity of the substrate material.

To quantify the impact of radioimpurities on the observed count rate, we conservatively  assume that the upper limits on the activities of \ce{^{226}Ra}, \ce{^{228}Th}, \ce{^{60}Co} and \ce{^{40}K} in the Si substrate (see~\cref{tab:gator_activities}) are positive detections. With a simple estimate, we obtain a rate of 0.4 events/day in the full \qty{88}{\milli\gram} \ce{Si} substrate. This number is a conservative upper limit on the event rate in the SNSPD, also because not every decay in the \ce{Si} substrate will lead to a count in the SNSPD. We thus conclude that these radioactivities do not have a significant contribution to the 15 events observed.
The oxygen-free copper sample holder was machined from spare low-activity material of the photomultiplier tube array support plates of the XENONnT DM experiment~\citesm{XENON:2024wpa}. 
A sample of this material with a mass of \qty{93.4}{\kilo\gram} has been screened at the GeMSE facility in the context of the XENONnT radiopurity control program~\citesm{Garcia:2022jdt, XENON:2021mrg}, with the resulting activities summarized in \cref{tab:gator_activities}. Auxiliary ICP-MS measurements were able to determine values of \qty{1.4(4)}{\micro\becquerel/\kilo\gram} and \qty{4(1)}{\micro\becquerel/\kilo\gram} for the specific activities of \ce{^238U} and \ce{^{228}Ra}, respectively.
We also note that the specific activities of the Cu holder are much lower than those of the Si substrate, and that the Cu holder is further away from the SNSPD than the substrate, leading to even lower count rates.

\bibliographystylesm{apsrev4-2}
\bibliographysm{references}

\end{document}